\documentclass[10pt,twocolumn,letterpaper]{article}

\usepackage[pagenumbers]{cvpr} 

\usepackage[dvipsnames]{xcolor}

\definecolor{cvprblue}{rgb}{0.21,0.49,0.74}
\usepackage[pagebackref,breaklinks,colorlinks,citecolor=cvprblue]{hyperref}

\def\paperID{*****} 
\def\confName{CVPR}
\def\confYear{2024}

\title{\Acs{2fa}: Navigating the Challenges and Solutions for Inclusive Access}

\author{Alexander Lengert\,\orcidlink{0000-0002-3884-9304}\\
Karlsruhe Institute of Technology (KIT)\\
Karlsruhe, Germany\\
{\tt\small alexander.lengert@student.kit.edu}
}

\usepackage{acro}
\usepackage{graphicx}
\usepackage{tabularx}
\usepackage{multirow}
\usepackage{subcaption}
\usepackage{orcidlink}

\newcommand{\enquote}[1]{``#1''}

\newcommand{\wcag}[1]{\acs{wcag}~#1}
\DeclareAcronym{ams}{
    short = AMS ,
    long = \emph{Authentication Method Selector}
}

\DeclareAcronym{captcha}{
    short = CAPTCHA ,
    long = \emph{Completely Automated Public Turing Test to Tell Computers and Humans Apart},
    long-plural-form = \emph{Completely Automated Public Turing Tests to Tell Computers and Humans Apart}
}

\DeclareAcronym{2fa}{
    short = 2FA ,
    long = \emph{Two-Factor Authentication}
}

\DeclareAcronym{ci}{
    short = CI ,
    long = \emph{Cognitive Impariment}
}

\DeclareAcronym{ecc}{
    short = ECC ,
    long = \emph{Elliptic Curve Cryptography}
}

\DeclareAcronym{fido}{
    short = FIDO ,
    long = \emph{Fast IDentity Onlines}
}

\DeclareAcronym{https}{
    short = HTTPS ,
    long = \emph{Hypertext Transfer Protocol Secure}
}

\DeclareAcronym{hi}{
    short = HI ,
    long = \emph{Hearing Impairment}
}

\DeclareAcronym{kit}{
    short = KIT ,
    long = \emph{Karlsruhe Institute of Technology}
}

\DeclareAcronym{mi}{
    short = MI ,
    long = \emph{Motor Impairment}
}

\DeclareAcronym{vi}{
    short = VI ,
    long = \emph{Visual Impairment}
}

\DeclareAcronym{ppg}{
    short = PPG ,
    long = Keystroke-induced \emph{Photoplethysmography}
}

\DeclareAcronym{qr_code}{
    short = QR Code ,
    long = \emph{Quick-Response Code}
}

\DeclareAcronym{sms}{
    short = SMS ,
    long = \emph{Short Message/Messaging Service}
}

\DeclareAcronym{sso}{
    short = SSO ,
    long = \emph{Single Sign-on}
}

\DeclareAcronym{totp}{
    short = TOTP ,
    long = \emph{Time-based One-time Password}
}

\DeclareAcronym{u2f}{
    short = U2F ,
    long = \emph{Universal 2nd Factor}
}

\DeclareAcronym{w3c}{
    short = W3C ,
    long  = \emph{World Wide Web Consortium},
    tag = included
}

\DeclareAcronym{wcag}{
    short = WCAG ,
    long  = \emph{Web Content Accessibility Guidelines}
}

\DeclareAcronym{who}{
    short = WHO ,
    long  = \emph{World Health Organization}
}

\DeclareAcronym{waiaria}{
    short = WAI-ARIA ,
    long  = \emph{Web Accessibility Initiative – Accessible Rich Internet Applications}
}

\newcommand{\footnotetextLink}[3]{\footnotetextLinkDate{#1}{#2}{#3}{n.d.}}
\newcommand{\footnotetextLinkDate}[4]{\footnotetext{#1. \textit{#2.} (#4). Fetched on 02/02/2024 via \href{https://#3}{\mbox{https://}#3}.}}

\newcommand{\pspace}[0]{\noindent{}}

\begin{document}
\maketitle
\begin{abstract}
\pspace{}The digital age requires strong security measures to protect online activities. \Ac{2fa} has emerged as a critical solution. However, its implementation presents significant challenges, particularly in terms of accessibility for people with disabilities. This paper examines the intricacies of deploying \ac{2fa} in a way that is secure and accessible to all users by outlining the concrete challenges for people who are affected by various types of impairments. This research investigates the implications of \ac{2fa} on digital inclusivity and proposes solutions to enhance accessibility. An analysis was conducted to examine the implementation and availability of various \ac{2fa} methods across popular online platforms. The results reveal a diverse landscape of authentication strategies. While \ac{2fa} significantly improves account security, its current adoption is hampered by inconsistencies across platforms and a lack of standardised, accessible options for users with disabilities. Future advancements in \ac{2fa} technologies, including but not limited to autofill capabilities and the adoption of \ac{fido} protocols, offer possible directions for more inclusive authentication mechanisms. However, ongoing research is necessary to address the evolving needs of users with disabilities and to mitigate new security challenges. This paper proposes a collaborative approach among stakeholders to ensure that security improvements do not compromise accessibility. It promotes a digital environment where security and inclusivity mutually reinforce each other.
\end{abstract}    
\section{Introduction}
\label{sec:intro}

As more and more aspects of our lives depend on the online world, the protection of online services is a critical issue in the digital age, in which we use digital platforms to access information, communicate and perform various tasks, including banking, education, healthcare and entertainment~\cite{kraus_2022}. However, these platforms are vulnerable to cyber-attacks that can compromise our personal data, privacy and identity. It is therefore vital to put strong security measures in place to protect our online accounts and transactions.

\acf{2fa} is an extensively implemented security measure that improves conventional login methods of a username and password by including an additional layer of authentication. This additional layer may include providing supplementary details like a code dispatched to the user's phone or email, a biometric scan, or a physical token~\cite{schneier_2005}. Recognised for its efficacy in mitigating the hazards of account takeover, phishing, and identity theft, \ac{2fa} is deemed an imperative component of online account security~\cite{oren_2022}.

Nevertheless, its adoption poses obstacles to digital accessibility, specifically for people with disabilities~\cite{andrew_2023, jain_2019}. Digital inclusivity is crucial in developing online platforms that serve users of all abilities, preferences, and circumstances. The pursuit of equal participation in the modern digital environment is not just a matter of convenience, it is a fundamental human right and a matter of obligation for society.

This paper aims to explore the digital accessibility challenges of implementing \ac{2fa} and propose solutions to overcome them. The paper addresses the question of whether cybersecurity should compromise accessibility, and how to strike a balance between security and usability. Additionally, the paper is intended to raise awareness and promote a culture of inclusivity among security practitioners, developers and users.

In \Cref{sec:understanding}, the crucial role of \ac{2fa} in online security is explored, emphasising its significance and the importance of digital accessibility. Proceeding to \Cref{sec:challenges}, the impact of disabilities on interacting with \ac{2fa} is assessed, highlighting the issues faced by diverse user groups with common \ac{2fa} approaches. \Cref{sec:choice} systematically explores various \ac{2fa} options, analysing which approaches are most common and how users with disabilities might confront hurdles when interacting with \ac{2fa} implementations.

\Cref{sec:future} delves into emerging trends in \ac{2fa} and proposes methods to enhance accessibility, anticipating advancements and addressing evolving user needs. Following this, \Cref{sec:discussion} discusses these findings, considering the broader implications for accessibility and security in digital environments.

The paper concludes in \Cref{sec:conclusion} by summarising key findings, recommending improvements in \ac{2fa} accessibility, and highlighting ongoing research avenues. The conclusion underscores the need for a sustained commitment to inclusive \ac{2fa} design and implementation.

Lastly, \Cref{sec:outlook} provides an outlook on future directions in \ac{2fa}, considering technological advancements, user behaviour trends, and regulatory changes. Actionable strategies for stakeholders to prepare for these future developments are proposed, emphasising the importance of early adoption of inclusive practices.

\section{Understanding \acs{2fa} and Accessibility}
\label{sec:understanding}

Accessibility involves the development of products, services, and digital platforms that are inclusive for all, including those with disabilities. In the digital sphere, this necessitates creating platforms with accessibility at the forefront and following established criteria that offer guidance for designers and developers. These standards secure the integration of accessibility into the design procedure from the beginning and provide benchmarks for assessing compliance, ensuring impartial access to digital platforms for all users~\cite{scullion_2021}.

The incorporation of \ac{2fa}, which mandates users to provide two distinct authentication factors, may pose potential challenges to accessibility. Certain \ac{2fa} methods may not be appropriate for individuals with disabilities~\cite{andrew_2023, jain_2019}. Therefore, it is essential to carefully evaluate the accessibility implications when integrating \ac{2fa} into digital platforms.

\subsection{Two-Factor Authentication (\acs{2fa})}
\label{sec:understanding:2fa}

It's common for the security benefits and threat models associated with implementing \ac{https} and \ac{2fa} to be misunderstood by end-users. They will often confuse encryption with authentication, downplay the security benefits of \ac{https}, and neglect and distrust its security indicators. These misconceptions can result in insecure configurations and usage behaviour~\cite{krombholz_2019}. \Ac{2fa} is actually intended to provide an additional layer of security beyond username and password. Requiring a second factor for authentication makes it increasingly arduous for unauthorised users to access an account, despite possessing the password~\cite{schneier_2005}.

The primarily used factors of multi-factor authentication, of which \ac{2fa} is a part, are~\cite{abhishek_2013}:

\begin{itemize}
    \item \textbf{Knowledge factor} (\enquote{Something you \textit{know}} -- such as a password or personal identification number),
    \item \textbf{Ownership factor} (\enquote{Something you \textit{have}} -- such as a physical token or smart card), 
    \item \textbf{Inherence factor} (\enquote{Something you \textit{are}} -- such as biometric techniques like fingerprint scanning or facial recognition), and
    \item \textbf{Social factor} (\enquote{\textit{Somewhere} you are or \textit{someone} you know} -- such as location-based or trust-based authentication methods).
\end{itemize}

\pspace{}Whilst these categories, or rather the use of more than of the factors in these categories, offer a more robust means of remote authentication than the conventional single-factor username and password approach, the implementation of \ac{2fa} also brings about a set of challenges, which may play into the fact that \ac{2fa} still isn't widely adopted by the general public~\cite{farke_2020}. In order to improve the situation, it is crucial to tailor \ac{2fa} solutions to the needs of users. By evaluating user feedback and dynamically adapting security policies, security mechanisms can be optimised from a user experience perspective. This method guarantees that \ac{2fa} is not only secure but also accessible, thereby encouraging more widespread use~\cite{feth_2015}, not only by people with disabilities.

\subsection{Accessibility Norms}
\label{sec:understanding:accessibility}

The \ac{w3c} has developed guidelines for the implementation of websites with the aim of ensuring accessibility. On the one hand, the \ac{wcag} guidelines help developers to create web content that is accessible to people with limited abilities. The provision of content can be tested and evaluated for compatibility at various levels. \Ac{wcag} refers to websites, which also includes the display of websites on mobile devices~\cite{wcag_2023}. 

The guidelines and success criteria for accessing and using web content from \ac{wcag} are based on four principles~\cite{wcag_2023}: 

\begin{enumerate}
    \item \textbf{Perceivability,}
    \item \textbf{Operability,}
    \item \textbf{Understandability} and
    \item \textbf{Robustness.}
\end{enumerate}

\pspace{}These principles are crucial for users with disabilities to be able to use the web. They ensure that information and user interface components are presentable to all users in a way that they can perceive, that user interface components and navigation are operable, that both information and user interface operation are understandable, and that content is robust enough to be reliably interpreted by a variety of user agents, including assistive technologies~\cite{wcag_2023}. \wcag{2.1}, published in 2018, extends \wcag{2.0}, published in 2008, by 17 additional criteria, including in the area of mobile accessibility and for people with \ac{vi}~\cite{wcag_2008, wcag_2018}.

\pspace{}Two new success criteria for improving authentication accessibility have been introduced by \wcag{2.2}~\cite{wcag_2023}:

\begin{itemize}
    \item \textbf{3.3.8 Accessible Authentication (Minimum/AA):} At least one authentication method must be available that does not require memorisation, transcription, or cognitive testing. This could involve allowing users to copy and paste passwords or use password managers.
    \item \textbf{3.3.9 Accessible Authentication (Enhanced/AAA):}  This standard is highly restrictive, mandating that all authentication methods be accessible and prohibiting the use of \acp{captcha}.
\end{itemize}

\pspace{}The criteria assume a major importance as they guarantee inclusivity of the authentication procedure for all individuals, including those with some kind of impairment. They also promote the use of more secure authentication methods, such as device-based authentication.

Furthermore, there is \ac{waiaria}, a specification of \ac{w3c} that helps developers to create web content that is accessible to people with limited abilities. \Ac{waiaria} defines specific attributes that can be added to HTML elements to provide a better experience for users with limited abilities~\cite{waiaria_2017}.

\subsection{Disability Types}
\label{sec:understanding:disabilities}

\citeauthor{crow_2008} differenciates between four types of impairments~\cite{crow_2008}. Their key characteristics as well as possible ways how can they influence how people affected by them interact with digital services will be discussed in this subsection.

\subsubsection{Visual Impairment}
\label{sec:understanding:disabilities:vi}

\Ac{vi} is one of the most common disabilities. This category includes blindness and all other types of low vision~\cite{congdon_2003}. The global number of individuals with \ac{vi} is estimated to be around 285 million, with 39 million people being blind and 246 million people with low vision, assuming a tolerance threshold of 20\% in both directions~\cite{pascolini_2012}.

\Ac{vi} can arise from a variety of causes. Refractive errors, where the length of the eyeball does not match the length of the eye's optical system; age-related causes; infectious causes (e.g. conjunctivitis, onchocerciasis and human immunodeficiency virus); nutritional and metabolic causes (e.g. vitamin A deficiency and diabetes); and all types of injury (e.g. work-related or from firearms) as possible triggers for \ac{vi}~\cite{congdon_2003}.

The technologies employed to alleviate \ac{vi} can be divided into two categories: audio-based technologies and visual-based technologies. The former refers to the use of screen readers, select-to-speak options, voice inputs, or even voice assistants. The latter pertains to the use of digital magnifying glasses, contrast enhancement, or colour inversion, among others~\cite{senjam_2021}.

\subsubsection{Hearing Impairment}
\label{sec:understanding:disabilities:hi}

According to the \ac{who}, approximately 430 million individuals worldwide, equivalent to around 5\% of the world's population, are in need of special treatment due to disabling hearing loss. The problem is expected to get worse, with estimates suggesting that more than 700 million people will be affected by 2050. Disabling hearing loss, exceeding 35 decibels, is particularly common in low- and middle-income countries, and its incidence usually increases with advancing age. Inherited factors and various environmental elements are among the diverse factors responsible for hearing loss. \Ac{hi}, varying in severity from mild to profound, has extensive implications for communication, cognition, education, and employment. The issue is exacerbated if left untreated and can lead to social exclusion~\cite{who_hi_2023}.

The impact of \ac{hi} on digital service use is significant. People with \ac{hi} often face challenges in accessing and using digital services due to the heavy reliance on audio cues on many digital platforms. As a result, people with \ac{hi} can feel isolated, frustrated and unable to access important information and services~\cite{arnaizsanchez_2023}.

From a digital accessibility standpoint, there are various strategies that can be employed to enhance the user experience for people with \ac{hi}. Subtitles or transcripts can be utilised as visual alternatives to audio content to enable people with \ac{hi} to access the same information as those without~\cite{hong_2010, hong_2011}. Secondly, the inclusion of sign language interpretation in video content can offer significant benefits, especially for those who primarily use sign language as a form of communication~\cite{debevc_2011}. Additionally, the implementation of haptic feedback, which utilises touch to communicate information, can be a valuable tool. For instance, vibrations or other tactile signals can notify users of crucial notifications or events~\cite{nanayakkara_2013}.

\subsubsection{Motor Impairment}
\label{sec:understanding:disabilities:mi}

\Ac{mi} covers a wide spectrum of conditions that affect a person's ability to move or control their muscles. According to the \ac{who}, an estimated 2.24 billion individuals worldwide, or about one in six people, live with a disabling motor disability. These limitations vary from mild to severe and present in diverse manners, causing difficulty in fine, gross motor functions or in a combination of both. The impact of \ac{mi} can be profound and may also affect an individual's usage of digital devices or access to online content~\cite{who_mi_2023}.

In the digital context, \ac{mi} can be a major obstacle, as traditional input methods such as mouse and keyboard interactions can be difficult or impossible for many people. A range of accessibility strategies can be employed to tackle these challenges~\cite{fanucci_2011}. One method is to allow alternate input methods besides the conventional mouse and keyboard interactions. Voice commands, head tracking, and eye gaze management, e.g., can empower people with limited fine motor abilities to navigate and interact with digital interfaces proficiently~\cite{trewin_2009}.

Moreover, enhancing the digital interfaces' accessibility can further improve the experience for people with \ac{mi}. This may require ensuring precise labeling for menus and buttons, properly sizing and spacing interactive elements, and employing accessible color schemes that don't fatigue visual perception. Moreover, integrating keyboard and accessibility shortcuts can help streamline navigation and enhance interaction for users with limited dexterity or motor control~\cite{naftali_2014}.

\subsubsection{Cognitive Impairment}
\label{sec:understanding:disabilities:ci}

\Ac{ci} comprises a spectrum of conditions that limit an individual's capacity to process, comprehend and retain information. These limitations can result in a variety of difficulties such as attention deficits, memory loss, language processing issues, and difficulties in executive functioning~\cite{britto_2021}. Such impairments can compromise an individual's ability to learn, comprehend and effectively engage with digital content~\cite{gordon_2005}. The exact figure of people with \ac{ci} around the world is difficult to determine due to a lack of standardised definitions and assessment methods~\cite{pais_2020}.

To tackle the concerns experienced by individuals with \ac{ci} in the digital landscape, several accessibility strategies can be employed. A fundamental approach is to simplify and improve the readability of digital content. This approach may entail utilising simple language with concise and coherent sentence structures~\cite{poulson_2004}, eschewing technical jargon and needlessly elaborate terminology, and offering alternate text descriptions for graphics and illustrations. Additionally, dividing intricate data into smaller, more digestible portions can promote comprehension and retention~\cite{friedman_2008}.

To further enhance accessibility for individuals with \ac{ci}, it is also beneficial to provide consistent and easy-to-follow navigation cues~\cite{leporini_2004}. Designing user interfaces involves implementing intuitive visual hierarchies, clear labelling for menus and headings, and organizing content logically and linearly. Furthermore, integrating visual cues and animations can assist users in navigating through complicated interfaces and improving their experience~\cite{moreno_2021}.

Lastly, deliberating cognitive load and accessibility features can further enhance the user experience for individuals with \ac{ci}. This can entail reducing the amount of superfluous content and elements on a page, providing unambiguous and uniform feedback for actions carried out, and proposing alternative methods to access information or accomplish tasks~\cite{lewis_2011}.

\section{Challenges in Implementing \acs{2fa} for People with Disabilities}
\label{sec:challenges}

\begin{table}[]
\centering
\begin{tabular}{l|c|c|c|c|}
\cline{2-5}
\multicolumn{1}{c|}{\textbf{}}                   & \textbf{\acs{vi}} & \textbf{\acs{hi}} & \textbf{\acs{mi}} & \textbf{\ac{ci}} \\ \hline
\multicolumn{1}{|l|}{\textbf{Perceivability}}    & $\times$    & $\times$    & $\times$    & $-$         \\ \hline
\multicolumn{1}{|l|}{\textbf{Operability}}       & $\times$    & $\times$    & $\times$    & $\times$    \\ \hline
\multicolumn{1}{|l|}{\textbf{Understandability}} & $-$         & $-$         & $-$         & $\times$    \\ \hline
\multicolumn{1}{|l|}{\textbf{Robustness}}        & $\times$    & $\times$    & $\times$    & $\times$    \\ \hline
\end{tabular}
\caption{Graphical summary of the results of \Cref{sec:challenges}: A cross~($\times$) indicates that \ac{2fa} methods may present a challenge to a person with a particular type of impairment (see \Cref{sec:understanding:disabilities}) in a particular accessibility principle, whereas a hyphen~($-$) indicates no impact.}
\label{tab:challenges}
\end{table}

This section is about identifying and analysing the challenges that people with different types of disabilities face when using \ac{2fa} methods for online authentication. The challenges have been categorised according to the four principles of web accessibility~(see \Cref{sec:understanding:accessibility}). A graphical summary of the results is provided in \Cref{tab:challenges}, which shows how each type of impairment affects the accessibility of \ac{2fa} methods in each principle.

\subsection{Perceivability}
\label{sec:challenges:perceivability}

Individuals with \ac{vi} face challenges in perceiving visual information on screens, such as \acp{captcha} or \acp{qr_code} in traditional \ac{2fa}. To improve perceivability, knowledge-based factors in \ac{2fa} should incorporate auditory or tactile cues, reducing reliance on visual cues~\cite{chen_2015,konoth_2020}. Similarly, non-visual sensory modalities, such as touch or sound, can enhance the verification of the inherence factor, which often depends on visual cues~\cite{konoth_2020,chen_2015}. The ownership factor in \ac{2fa} can also benefit from devices that provide tactile or auditory feedback, such as hardware tokens or biometric devices, improving the accessibility for users with \ac{vi}~\cite{chen_2015}.

For individuals with \ac{hi}, auditory cues in \ac{2fa} models, like phone call verifications or voice-based instructions, present challenges~\cite{zhu_2022}. Proposed solutions include incorporating visual cues, such as flashing lights or colour-coded indicators, to enhance accessibility for users with \ac{hi}~\cite{ward_2019}.

\Ac{2fa} models relying on memory abilities, such as \acf{totp} tokens or \acf{sms} techniques, pose challenges for individuals with \ac{ci}. Alternative approaches, such as wearable technology automation, can address these challenges~\cite{chen_2015}.

\subsection{Operability}
\label{sec:challenges:operability}

Individuals with \ac{vi} may face difficulties with manual transfer between devices or applications in common \ac{2fa} models. Automating the \ac{2fa} process through wearables may enhance operability for \ac{vi} users, streamlining their interaction with online services~\cite{chen_2015}.

Individuals with disabilities encounter challenges beyond perceptual issues with traditional \ac{2fa} methods. For \ac{mi} and \ac{vi} individuals, especially those with poor or no eyesight relying heavily on tactile feedback, tasks involving physical actions such as pressing buttons or entering codes pose significant challenges~\cite{kumar_2019}.

Operability challenges manifest for individuals with \ac{mi} in their limited ability to interact with and navigate devices or systems, essential for traditional \ac{2fa} methods. Fine motor skills and precise movements required for authentication processes create barriers to successful completion~\cite{rajavenkatanarayanan_2019}. Limited mobility and alternative input methods make understanding and interacting with traditional \ac{2fa} methods challenging for people with \ac{mi}. Clinical tests assessing \ac{mi} often require direct interaction with healthcare professionals, limiting independence in utilising traditional \ac{2fa} methods~\cite{rajavenkatanarayanan_2019}.

Cognitive accessibility challenges impact \ac{2fa} operability for people with \ac{ci}. Specific cognitive disabilities, including memory, attention, concentration, and problem-solving skills, are crucial for successful operability~\cite{erinola_2023}. \Ac{ci} significantly impacts the understanding and adoption of \ac{2fa}. Attentional limitations, particularly processing speed, become critical for engaging with security measures like \ac{2fa}~\cite{moreno_2023,neupert_2021}. Considering cognitive accessibility needs when designing \ac{2fa} interfaces is crucial, involving design patterns and guidelines to enhance usability for individuals with \ac{ci}~\cite{moreno_2023}.

\subsection{Understandability}
\label{sec:challenges:understandability}

Designing knowledge-based factors that are simple, intuitive, and easy to understand is crucial for \ac{vi} users. Complex or confusing instructions associated with \acs{sms}-based techniques create significant barriers. Simplifying the \ac{2fa} process is essential to enhance overall user experience~\cite{chen_2015}.

Individuals with \ac{ci} may face challenges in comprehending and memorising complex authentication information, such as passwords or PINs~\cite{erinola_2023}.

Design considerations, such as simplifying language, providing clear instructions, minimising information overload, and ensuring appropriate line spacing, can improve understandability for individuals with \ac{ci}. Glossary mechanisms or tools detecting complex words and offering simpler replacements or definitions during the authentication process can improve comprehension for individuals with \ac{ci}~\cite{moreno_2023}.

\subsection{Robustness}
\label{sec:challenges:robustness}

Existing \ac{2fa} systems may incorporate accessibility features, but the effectiveness of these features requires further evaluation. Different evaluation tools yield varying results, indicating discrepancies in implementation and effectiveness~\cite{ismailova_2022}.

Exploring stronger authentication protocols, coupled with knowledge-based factors, to increase the robustness of \ac{2fa} methods is an area yet to be thoroughly examined. Balancing the additional security layer provided by knowledge-based factors with usability is crucial for individuals with \ac{vi}~\cite{chen_2015}.

Traditional \ac{2fa} methods may lack security for individuals with \ac{mi}, given limited mobility or alternative input methods. Securely interacting with authentication mechanisms through assistive devices and inputting passwords or providing biometric samples pose challenges for individuals with \ac{mi}~\cite{kumar_2019}.

Social factors, such as support from caregivers or peers, play a vital role in influencing the adoption and acceptance of \ac{2fa} by people with \ac{ci}. Caregiver or peer assistance shapes the authentication experience, contributing to the overall robustness of the \ac{2fa} system~\cite{erinola_2023}.

To enhance robustness, creating \ac{2fa} solutions that are accessible and user-friendly, specifically tailored to individuals with \ac{ci}, is essential. Considering social dynamics and support structures is crucial in this context~\cite{erinola_2023}.

\section{Choice and Availability}
\label{sec:choice}

\begin{figure*}
    \centering
    \begin{subfigure}{\linewidth}
        \centering
        \begin{tabular}{ccc}
            \multicolumn{2}{c}{\begin{subfigure}{0.45\linewidth}
                                    \centering
                                    \includegraphics[width=\linewidth]{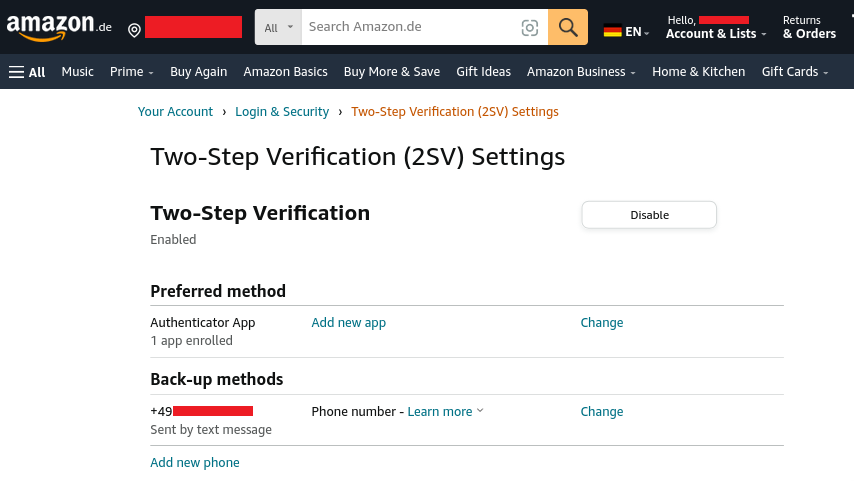}
                                    \caption{}\label{fig:2fa_amazon}
                                \end{subfigure}} & 
            \multirow{2}{*}[3.55cm]{\begin{subfigure}{0.275\linewidth}
                                        \centering
                                        \includegraphics[width=\linewidth]{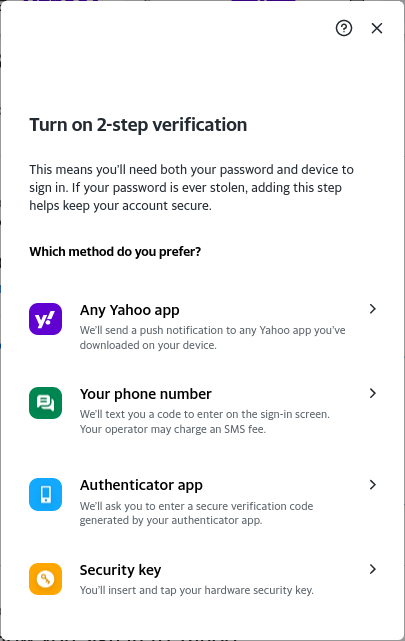}
                                        \caption{}\label{fig:2fa_yahoo}
                                   \end{subfigure}} \\
            \begin{subfigure}{0.3\linewidth}
                \centering
                \includegraphics[width=\linewidth]{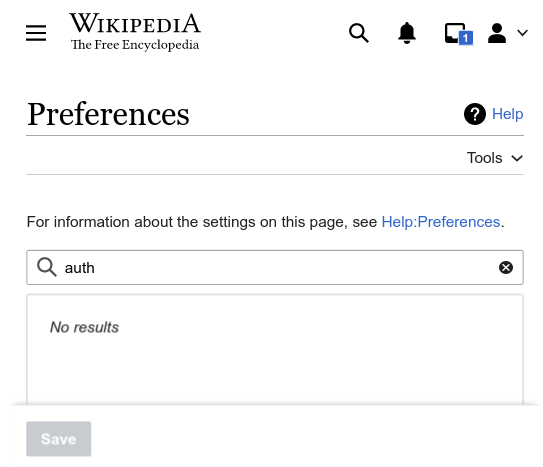}
                \caption{}\label{fig:2fa_wikipedia}
            \end{subfigure} & 
            \begin{subfigure}{0.2675\linewidth}
                \centering
late                \includegraphics[width=\linewidth]{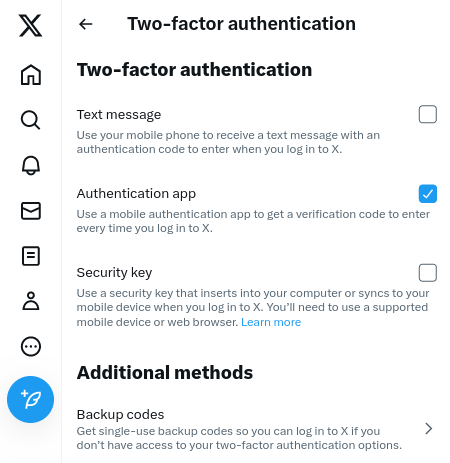}
                \caption{}\label{fig:2fa_x}
            \end{subfigure} & {}
        \end{tabular}
    \end{subfigure}
    \caption{Examples of \acs{2fa} settings of some popular websites on the internet: (a) Amazon's \acs{2fa} settings contain \acs{totp} without backup codes and SMS. (b) Yahoo's \acs{2fa} settings contain Push, \acs{sms}, \acs{totp} with backup codes, and \acs{u2f}/\acs{fido}. (c) Wikipedia's \acs{2fa} settings are not available with a normal user account. (d) X's \acs{2fa} settings contain \acs{sms}, \acs{totp} with backup codes, and \acs{u2f}/\acs{fido}.}
    \label{fig:2fa-settings}
\end{figure*}

One of the goals of this paper is to investigate the choice of \ac{2fa} technologies that end users have when they use online services. To answer this question, an inspection of the ten most popular websites on the internet, excluding solely pornographic websites and websites that do not offer the creation of an account (e.g. DuckDuckGo). We also included the \ac{kit} \ac{sso} as a case of special interest due to the fact that this paper's target audience is mostly associated with that university. Therefore, they should be familiar with the \ac{sso} system and should be able to widen the choice of \ac{2fa} methods. 

\subsection{Exploring \acs{2fa} Methods Across Platforms}
\label{sec:choice:exploring}

For each website, it was checked which \ac{2fa} methods they supported. The five \ac{2fa} methods present on the examined web services are the most common ones on the internet~\cite{reese_2019}. Namely they are:

\begin{enumerate}
    \item \acf{sms}: Ownership Factor (you \textit{own} the SIM card that is connected to the phone number the \ac{sms} gets sent to),
    \item \acf{totp}: Ownership Factor (you \textit{own} the device \ac{totp} is set up on),
    \item Backup Codes/\enquote{Codes}: (Knowledge Factor: you \textit{know} the backup codes given to you when you set up \ac{totp}),
    \item Push: Ownership Factor (you \textit{own} the phone where the app you get the notification on is installed on), and
    \item \acf{u2f}/\acf{fido}: Ownership Factor (you \textit{own} the key).
\end{enumerate}

\pspace{}By taking a look at \Cref{tab:availability}, where the recorded results as well as the observations are summarised graphically, it becomes apparent that the most common \ac{2fa} methods are \ac{sms}, \ac{totp}, and codes. All of the websites, except Wikipedia~(see \Cref{fig:2fa_wikipedia}) and \ac{kit} \ac{sso}, support \ac{sms} as a \ac{2fa} method. \ac{totp} and codes are supported by nine and eight websites, respectively. Push notifications are supported by five websites, namely Google, YouTube, Facebook, Instagram, and Yahoo~(see \Cref{fig:2fa_yahoo}). \ac{u2f}/\ac{fido} devices are supported by four websites, namely Google, YouTube, Yahoo, and X (fka Twitter)~(see \Cref{fig:2fa_x}). Wikipedia and \ac{kit} \ac{sso} are the only examined web services that do not support any push notifications or \ac{u2f}/\ac{fido} devices. Moreover, Wikipedia only offers \ac{2fa} to certain user groups, such as administrators and editors. It is therefore the only major website analysed in this paper that doesn't offer \ac{2fa} support to all its users~(see \Cref{fig:2fa_wikipedia}).

\pspace{}Another observation from the conducted analysis is that some websites allow users to log in with their accounts from other websites, such as Google, Apple, or Meta (fka Facebook). This means that users can leverage the \ac{2fa} methods of those websites, even if the original website does not support them. For example, X (fka Twitter) does not support push notifications, but users can log in with their Google account, which does support push notifications. This also implies that users may have different \ac{2fa} options depending on which account they use to log in.

\subsection{Observations and Implications}
\label{sec:choice:observations}

In summary, there is a variety of \ac{2fa} methods available to end users when using online services, but not all websites support the same methods. Users may also have different \ac{2fa} options depending on which account they use to log in. These findings suggest that there is no universal standard for \ac{2fa} implementation, and that users may face different levels of security and usability when they use \ac{2fa} on different websites. 

Microsoft and Yahoo offer all the different \ac{2fa} methods that appear in \Cref{tab:availability}~(see \Cref{fig:2fa_yahoo}), whereas most websites exclude some form or another from their available \ac{2fa} schemes. In the rather extreme case of Wikipedia, the common user doesn't get the possibility to secure his/her account with a second factor at all~(see \Cref{fig:2fa_wikipedia}), which, of course, has a negative influence on most user accounts' security. On the other hand, the issue of \ac{2fa} accessibility for people with disabilities is completely avoided when there is no \ac{2fa} in the first place.

The lack of choice especially affects people with disabilities as a wider variety of options would facilitate better accessibility to a more secure way to use online services, as people with different kinds of impairments could choose the \ac{2fa} method that suits their needs in the best way~(see \Cref{sec:challenges}).

\section{Future Trends and Improvements in Accessible \acs{2fa}}
\label{sec:future}

Acknowledging the complexities and requirements within the sphere of \ac{2fa} for individuals with disabilities is crucial for the creation of more inclusive systems~(see \Cref{sec:challenges}). The present state of affairs indicates a substantial gap in research concerning the comprehensive accessibility of authentication methods for users with disabilities. It is evident that none of the widely adopted methods completely address the needs of this diverse demographic, highlighting the critical need for innovation and enhancement in \ac{2fa} technologies~\cite{erinola_2023}.

\subsection{Advancements in \acs{2fa} Technologies}
\label{sec:future:advancements}

Exploring the details, the P2Auth system emerges as a notable advancement. This system amalgamates PIN entry with \ac{ppg} measurements, thereby heightening security against vulnerabilities such as shoulder surfing attacks and the exploitation of weak passwords. This methodology is particularly advantageous for individuals with \ac{vi}, relying less on visual interaction and thus augmenting perceivability. Moreover, it accommodates those with \ac{mi} by offering support for both single-handed and dual-handed input modes, improving operability. The combination of knowledge-based (PIN) and inherence-based (biometric PPG measurements) factors in P2Auth solidifies its standing as a sophisticated, accessible approach to \ac{2fa}~\cite{su_2023}.

The discourse on biometric systems for cognitive accessibility directs attention to users with \ac{ci}. Conventional methods like passwords and PINs are challenging for individuals with dyslexia or dyscalculia, given their reliance on character recognition and numerical computation. In contrast, biometric systems, which include technologies such as fingerprint scanning or voice recognition, present a more accessible alternative. These systems, aligning with the inherence factor, leverage unique physical or behavioural characteristics, thereby diminishing cognitive strain and enhancing user-friendliness~\cite{dicampi_2023}.

\pspace{}For users with \ac{vi}, the research by \citeauthor{dosono_2015} sheds light on considerable difficulties in locating authentication interfaces and interacting with assistive technologies like screen readers. Although this study does not propose a specific technological solution, it emphasises the necessity for more inclusively designed authentication systems, underlining the significance of perceivability and understandability in the design of \ac{2fa} systems~\cite{dosono_2015}.

In the realm of cloud computing, the attribute-based access control system delineated by \citeauthor{liu_2015} means a significant advancement in security and privacy. This system is particularly effective for users with \ac{ci}, as it alleviates the mental burden associated with remembering complex passwords or navigating convoluted authentication procedures, thus enhancing operability. The system integrates the knowledge factor (user's secret key) with the ownership factor (security device), offering a multi-layered security strategy~\cite{liu_2015}.

The framework tailored for cloud computing environments merges conventional user ID, password, and One-Time Password (OTP) verification processes with sophisticated cryptographic methods. This framework is notably beneficial for individuals with \ac{ci} and \ac{mi}. The employment of OTPs and straightforward password criteria lessens cognitive demands, while the reliance on standard input techniques renders it accessible to those with \ac{mi}. This system exemplifies the integration of knowledge and ownership factors in \ac{2fa}~\cite{kaur_2022}.

Turning to mobile platforms, the two-factor authentication system that incorporates handwritten signatures alongside passwords, as introduced by \citeauthor{khan_2015}, presents a unique combination of security and user convenience. The utilisation of handwritten signatures addresses the inherence factor and is particularly accessible to those with certain \ac{mi} who may find conventional keyboard or mouse inputs challenging~\cite{khan_2015}.

The multi-factor authentication system featuring the \ac{ams} technique introduces adaptability in selecting authentication methods based on user behaviour. This system is advantageous for individuals with cognitive challenges, allowing for simpler and more familiar authentication procedures. By analysing user behaviour, including location and web browser usage, this framework adds an additional layer of security and caters to the knowledge, inherence, and ownership factors~\cite{mostafa_2023}.

The examination of web services' security concerns by \citeauthor{zezulak_2023} draws attention to the issues faced by disabled users, particularly in authentication processes like \ac{captcha} completion. Suggested solutions such as passtones and Spoken \ac{captcha}  improve accessibility for users with various disabilities, that may turn out to be useful especially in regard of the knowledge and inherence factors~\cite{zezulak_2023}.

{\begin{table*}
  \centering
  \begin{subtable}{\linewidth}
    \small
\begin{tabularx}{\linewidth}{|p{1.995cm}|*{5}{>{\centering\arraybackslash}X|}p{5.545cm}|}
  \cline{2-7}
  \multicolumn{1}{c|}{} & \textbf{\acs{sms}} & \textbf{\acs{totp}} & \textbf{Codes} & \textbf{Push} & \textbf{\acs{u2f}/\acs{fido}} & \textbf{Observations} \\ \hline
  \textbf{Google}       & $\checkmark$ & $\checkmark$  & $\checkmark$            & $\checkmark$  & $\checkmark$      &                       \\ \hline
  \textbf{YouTube}      & $\checkmark$ & $\checkmark$  & $\checkmark$            & $\checkmark$  & $\checkmark$      & Uses Google acct.    \\ \hline
  \textbf{Facebook}     & $\checkmark$ & $\checkmark$  & $\checkmark$            & $\checkmark$  & --                & Uses Meta (fka Facebook) acct.                      \\ \hline
  \textbf{X (fka Twitter)} & $\checkmark$ & $\checkmark$  & $\checkmark$            & --            & $\checkmark$      & Allows log-in with Apple and Google accts.                             \\ \hline
  \textbf{Wikipedia}    & --            & ($\checkmark$)  & ($\checkmark$)            & --            & --                & Only certain user groups have access to \ac{2fa} \\ \hline
  \textbf{Instagram}    & $\checkmark$ & $\checkmark$  & $\checkmark$            & $\checkmark$  & --                & Uses Meta (fka Facebook) acct.                                                     \\ \hline
  \textbf{Reddit}       & --            & $\checkmark$  & $\checkmark$            & --            & --                & Allows log-in with Apple and Google accts.                             \\ \hline
  \textbf{Amazon}       & $\checkmark$ & $\checkmark$  & --                        & --            & --                &                       \\ \hline
  \textbf{Yahoo}        & $\checkmark$ & $\checkmark$  & $\checkmark$            & $\checkmark$  & $\checkmark$      & Allows log-in with Google acct.                                        \\ \hline
  \textbf{Bing}         & $\checkmark$ & $\checkmark$  & $\checkmark$            & $\checkmark$  & $\checkmark$      & Uses Microsoft acct.                                                    \\ \hline
  \textbf{...}          & ...          & ...           &  ...                    & ...           & ...               & \\ \hline
  \textbf{KIT \acs{sso}}      & --            & $\checkmark$  & $\checkmark$            & --            & $\checkmark$      &                       \\ \hline
\end{tabularx}
  \end{subtable}
    \caption[Caption]{Availability of \ac{2fa} methods on the ten most popular websites on the internet\footnotemark{}, excluding solely pornographic websites and such websites that don't offer the creation of an account, including \acs{kit} \acf{sso}.}
    \label{tab:availability}
\end{table*}
\footnotetextLink{Semrush Inc}{Most Visited Websites in the World, December 2023}{www.semrush.com/website/top/}}

\pspace{}The robust Two-Factor Authentication protocol based on \ac{ecc}, as developed by \citeauthor{liu_2023}, marks another significant progression. This protocol increases accessibility for individuals with \ac{ci} and \ac{mi} by simplifying the authentication process. It addresses the knowledge factor by enhancing usability and incorporates the inherence factor through the integration of biometric data~\cite{liu_2023}.

The Easier Web System aims to assist individuals with \ac{ci} in comprehending textual content. Employing cognitive accessibility design principles, this system aids users with \ac{ci} by improving text readability and understandability~\cite{moreno_2021}.

The PiGy system utilises the piezo-gyro channel for transmitting OTPs, thereby obviating the need for manual input and benefiting users with \ac{vi} and \ac{mi}. This system exemplifies the ownership factor and bolsters operability~\cite{oren_2022}.

\subsection{Enhancing Accessibility and Security}
\label{sec:future:enhancing}

\ac{totp} are a widely adopted form of  \ac{2fa}. It generates passwords that are valid for a brief period, enhancing security by requiring something the user has, e.g., a mobile device~(see \Cref{fig:totp-now}), and knows, e.g., a PIN or password~(see \Cref{sec:choice}). However, the conventional practice of manually transferring \acp{totp} from a separate application, such as a mobile app, to a login prompt introduces friction for users, especially ones with different types of disabilities~(see \Cref{sec:understanding:disabilities}), and potential for error~(see \Cref{sec:challenges}).

The introduction of \ac{totp} autofill capabilities aims to simplify the process of entering \acp{totp} by automatically inputting them~(see \Cref{fig:totp-autofill-demo}). This feature works across all common browsers and integrates with iOS and Android platforms. These advancements not only eliminate the need for manual code entry but also assist users with \ac{vi} or \ac{mi}, reducing the risks of incorrect code use. The functionality may also be beneficial for people with \ac{ci}, as once set up, the process of filling in \acp{totp} can be fully automated.

\subsection{Balancing Convenience with Security}
\label{sec:future:balancing}

However, the convenience of autofill functionality introduces privacy risks due to potential discrepancies between how browsers interpret web pages and what users can see. \citeauthor{lin_2020} highlighted the privacy threats of browser form autofill, revealing vulnerabilities that could be exploited to obtain sensitive user information stealthily~\cite{lin_2020}. This underscores the need for ongoing improvements in autofill technology to balance user convenience with security and privacy protections.

The integration of \ac{fido} protocols, WebAuthn~\cite{webauthn_2021} and the development of passkeys represents a significant step forward in making digital authentication more accessible and secure, particularly for users with disabilities.

FIDO2 provides a secure and user-friendly alternative to traditional password-driven logins through the use of biometrics and hardware security keys~\cite{webauthn_2021}. This approach is particularly beneficial for users with disabilities, as it is a vast simplification of the authentication process. For instance, individuals with \ac{vi} or \ac{mi} can authenticate through easier, more accessible means such as fingerprint scanning or a single touch of a hardware key~(see \Cref{fig:fido-prompt}), bypassing the need for complex passwords.

The widespread adoption of \ac{fido} by leading service providers~(see \Cref{tab:availability}) and its support across major browsers and operating systems highlights its role in providing secure web authentication and accessibility to a wider user base. It enables users to use familiar devices with built-in accessibility features, resulting in a seamless and inclusive authentication experience.

Passkeys further enhance digital accessibility by providing password-free logins using the biometric sensing capabilities of a device and seamless integration with the accessibility preferences of the user. They seamlessly integrate with users' preferred accessibility settings and sync across devices~(see \Cref{fig:passkey-setup}), ensuring users can access their accounts without repeated authentication steps, regardless of the device in use. This synchronisation increases security and promotes a consistent digital experience for users with disabilities by eliminating the need for different authentication methods on each device.

\section{Discussion}
\label{sec:discussion}

{\begin{figure}[t]
    \centering
    \includegraphics[width=0.75\linewidth]{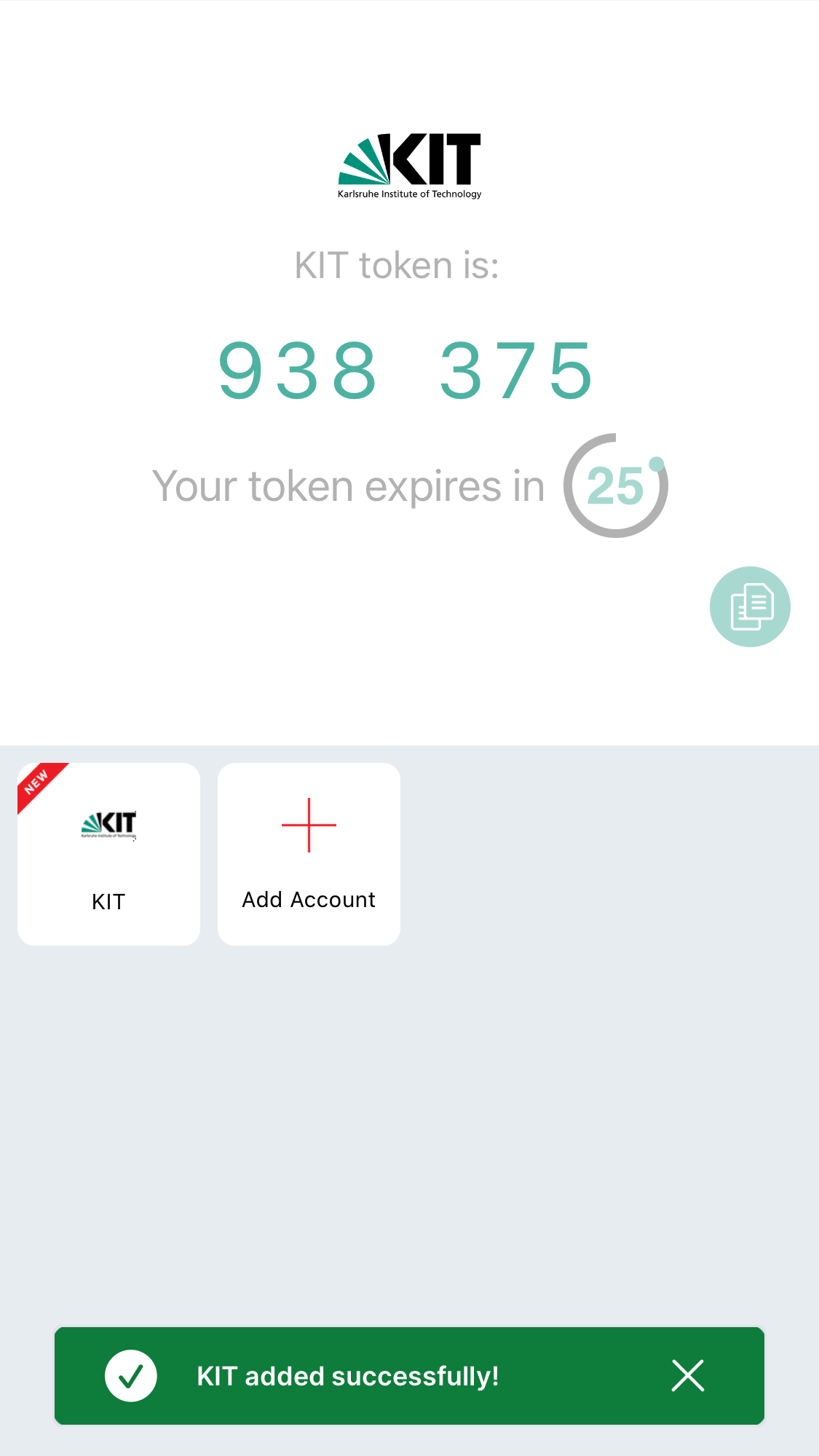}
    \caption[Caption]{\ac{totp} setup in Twilio Authy.\footnotemark{}}
    \label{fig:totp-now}
\end{figure}
\footnotetextLink{Twilio Inc}{
Welcome to Authy}{support.authy.com/hc/en-us/articles/115001943608-Welcome-to-Authy}}

{\begin{figure}[t]
    \centering
    \href{https://lngrt.de/totp-autofill-demo}{\includegraphics[width=0.75\linewidth]{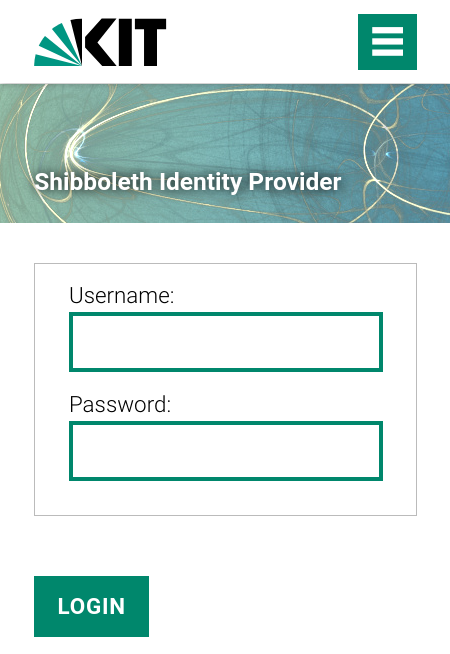}}
    \caption[Caption]{\Ac{totp} autofill functionality\footnotemark{} demo. \href{https://lngrt.de/totp-autofill-demo}{Show the animation.}}
    \label{fig:totp-autofill-demo}
\end{figure}
\footnotetextLink{Bitwarden, Inc}{Auto-fill Logins in Browser Extensions}{bitwarden.com/help/auto-fill-browser/\#totp-auto-fill}}

This paper presents an analysis of the implementation of \ac{2fa} for individuals with disabilities. The findings highlight the multifaceted nature of implementing \ac{2fa} in a way that is accessible to individuals with disabilities. The effectiveness of \ac{2fa} technologies is limited by their inconsistent and non-standardised implementation across different platforms~(see \Cref{sec:challenges}). Not only does this inconsistency complicate the user experience, but it also disproportionately affects people with disabilities, who may face additional challenges in navigating these systems~(see \Cref{sec:understanding:disabilities}).

There is a significant gap between the potential for improved security and the actual user experience, particularly for users with disabilities, when integrating \ac{2fa} methods into digital platforms~(see \Cref{sec:choice:observations}). The range of disabilities, including \ac{vi} and \ac{hi}, as well as \ac{mi} and \ac{ci}, emphasises the need for a holistic approach to \ac{2fa} design that prioritises accessibility and user autonomy. Relying on external assistance to establish \ac{2fa} methods undermines the autonomy and security these systems aim to provide, highlighting the need for more intuitive and standardised \ac{2fa} solutions.

Furthermore, analysis of the availability and implementation of different \ac{2fa} methods on popular online platforms reveals a patchwork of approaches, with some platforms offering extensive options and others lagging behind~(sec \Cref{sec:choice}). This disparity not only affects the overall security posture of users across different platforms but also limits the choices available to users with disabilities, who might benefit from a broader selection of accessible \ac{2fa} methods.

The examination of forthcoming developments and enhancements in \ac{2fa} technologies, such as the incorporation of \ac{totp} autofill capabilities and the implementation of \ac{fido} protocols, provides insight into the possibility of more convenient and user-friendly authentication mechanisms~(see \Cref{sec:future:enhancing}). Nevertheless, these advancements also reveal new obstacles in balancing convenience with security and privacy concerns, particularly with autofill vulnerabilities~(see \Cref{sec:future:balancing}).

\section{Conclusion}
\label{sec:conclusion}

This paper has endeavoured to bridge the gap between the need for robust online security measures and the imperative of digital accessibility, with a focus on the implementation of \ac{2fa} methods. Through a comprehensive examination of the challenges faced by users with disabilities, the analysis has highlighted the critical areas where current \ac{2fa} methods fall short in terms of accessibility and usability.

The discussion on future trends and improvements in \ac{2fa} technologies underscores the potential for creating more inclusive and user-friendly authentication systems. However, it also emphasises the importance of ongoing research and development to address the evolving needs of users with disabilities and to mitigate new security and privacy challenges that may arise from these technological advancements.

{\begin{figure}[t]
    \centering
    \includegraphics[width=0.85\linewidth]{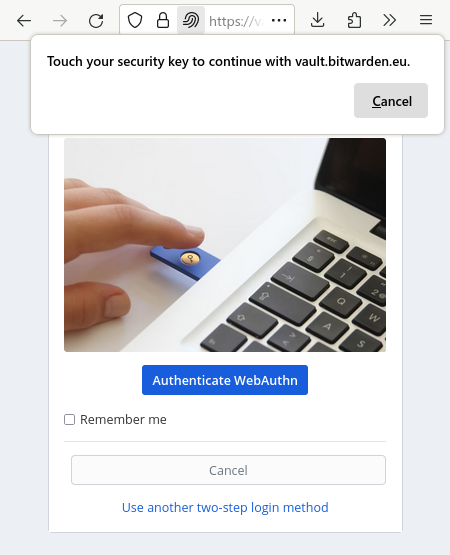}
    \caption[Caption]{Prompt for \ac{fido} authentication.\footnotemark{}}
    \label{fig:fido-prompt}
\end{figure}
\footnotetextLink{Bitwarden, Inc}{Two-step Login via FIDO2 WebAuthn}{bitwarden.com/help/setup-two-step-login-fido}}

{\begin{figure}[t]
    \centering
    \includegraphics[width=0.7905\linewidth]{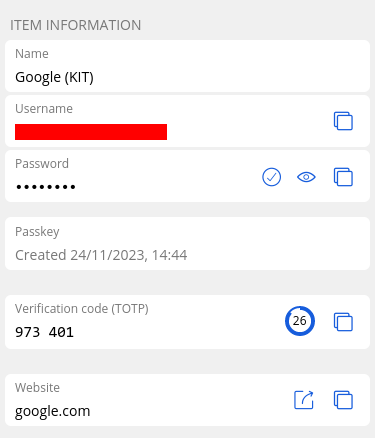}
    \caption[Caption]{Passkey set up in Bitwarden.\footnotemark{}}
    \label{fig:passkey-setup}
\end{figure}
\footnotetextLink{Bitwarden, Inc}{Storing Passkeys}{bitwarden.com/help/storing-passkeys/\#passkey-storage}}

In conclusion, although \ac{2fa} is crucial for online security, its current implementation presents significant accessibility challenges for users with disabilities. This paper's findings call for a collaborative effort among security practitioners, developers, and users to foster an inclusive culture and ensure that security measures do not compromise accessibility. By promoting collaboration and innovation in designing and implementing \ac{2fa} systems, we can move closer to a secure and accessible digital environment.

\section{Future Research}
\label{sec:outlook}

The exploration of accessible \ac{2fa} systems opens up several avenues for future research. One critical area involves the development of standardised guidelines for the implementation of \ac{2fa} methods that are universally accessible. Such guidelines would provide a framework for designing authentication systems that cater to the diverse needs of users with disabilities, ensuring that security measures enhance rather than hinder their online experience.

Additionally, further research is needed to evaluate the effectiveness and user acceptance of emerging \ac{2fa} technologies, such as \ac{fido} and Passkeys, among individuals with disabilities. Studies focusing on the real-world application of these technologies can offer valuable insights into their usability, security, and potential barriers to adoption.

Investigating the privacy implications of new \ac{2fa} methods, particularly in the context of autofill functionalities, represents another crucial research direction. Understanding the trade-offs between convenience and privacy will be essential in designing \ac{2fa} systems that protect user data while offering an accessible authentication experience.

Lastly, interdisciplinary research that combines insights from cybersecurity, human-computer interaction, and disability studies can foster innovative solutions to the accessibility challenges of \ac{2fa}. By embracing a holistic approach to security and accessibility, future research can contribute to creating a more inclusive and secure digital landscape.

{
    \small
    \bibliographystyle{ieeenat_fullname}
    \bibliography{main}

\begin{thebibliography}{58}
\providecommand{\natexlab}[1]{#1}
\providecommand{\url}[1]{\texttt{#1}}
\expandafter\ifx\csname urlstyle\endcsname\relax
  \providecommand{\doi}[1]{doi: #1}\else
  \providecommand{\doi}{doi: \begingroup \urlstyle{rm}\Url}\fi

\bibitem[Abhishek et~al.(2013)Abhishek, Roshan, Kumar, and Ranjan]{abhishek_2013}
Kumar Abhishek, Sahana Roshan, Prabhat Kumar, and Rajeev Ranjan.
\newblock A comprehensive study on multifactor authentication schemes.
\newblock In \emph{Advances in Computing and Information Technology}, pages 561--568, Berlin, Heidelberg, 2013. Springer Berlin Heidelberg.

\bibitem[Andrew et~al.(2023)Andrew, Watson, Oh, and Tigwell]{andrew_2023}
Sarah Andrew, Stacey Watson, Tae Oh, and Garreth~W. Tigwell.
\newblock Authentication challenges in customer service settings experienced by deaf and hard of hearing people.
\newblock \emph{Extended Abstracts of the 2023 CHI Conference on Human Factors in Computing Systems}, 2023.

\bibitem[Arnaiz-S{\'a}nchez et~al.(2023)Arnaiz-S{\'a}nchez, Carriqu{\'i}, Alcaraz, and Caballero]{arnaizsanchez_2023}
Pilar Arnaiz-S{\'a}nchez, Patricia~Zorrilla Carriqu{\'i}, Salvador Alcaraz, and Carmen~Mar{\'i}a Caballero.
\newblock A study of the barriers to communication and learning of university students with hearing impairment during the covid-19 pandemic.
\newblock \emph{European Journal of Special Needs Education}, 38:\penalty0 928 -- 938, 2023.

\bibitem[Campi and Luccio(2023)]{dicampi_2023}
Alessia Michela~Di Campi and Flaminia Luccio.
\newblock Accessible authentication methods for people with diverse cognitive abilities, 2023.

\bibitem[Chen and Goh(2015)]{chen_2015}
Alex~Qiang Chen and Weihan Goh.
\newblock Two factor authentication made easy.
\newblock In \emph{International Conference on Web Engineering}, 2015.

\bibitem[Congdon et~al.(2003)Congdon, Friedman, and Lietman]{congdon_2003}
Nathan~G Congdon, David~S Friedman, and Thomas Lietman.
\newblock Important causes of visual impairment in the world today.
\newblock \emph{Jama}, 290\penalty0 (15):\penalty0 2057--2060, 2003.

\bibitem[Crow(2008)]{crow_2008}
Kevin~L. Crow.
\newblock Four types of disabilities: Their impact on online learning.
\newblock \emph{TechTrends}, 52:\penalty0 51--55, 2008.

\bibitem[Debevc et~al.(2011)Debevc, Kosec, and Holzinger]{debevc_2011}
Matja{\v{z}} Debevc, Primo{\v{z}} Kosec, and Andreas Holzinger.
\newblock Improving multimodal web accessibility for deaf people: sign language interpreter module.
\newblock \emph{Multimedia Tools and Applications}, 54:\penalty0 181--199, 2011.

\bibitem[Dosono et~al.(2015)Dosono, Hayes, and Wang]{dosono_2015}
Bryan Dosono, Jordan Hayes, and Yang Wang.
\newblock \enquote{I'm Stuck}: A contextual inquiry of people with visual impairments in authentication.
\newblock In \emph{Eleventh Symposium On Usable Privacy and Security (SOUPS 2015)}, pages 151--168, 2015.

\bibitem[Erinola et~al.(2023)Erinola, Buckmann, Friedauer, Yard{\i}m, and Sasse]{erinola_2023}
Ahmet Erinola, Annalina Buckmann, Jennifer Friedauer, Asl{\i} Yard{\i}m, and M~Angela Sasse.
\newblock “as usual, i needed assistance of a seeing person”: Experiences and challenges of people with disabilities and authentication methods.
\newblock In \emph{2023 IEEE European Symposium on Security and Privacy Workshops (EuroS\&PW)}, pages 575--593. IEEE, 2023.

\bibitem[Fanucci et~al.(2011)Fanucci, Iacopetti, and Roncella]{fanucci_2011}
L. Fanucci, F. Iacopetti, and R. Roncella.
\newblock A console interface for game accessibility to people with motor impairments.
\newblock In \emph{2011 IEEE International Conference on Consumer Electronics -Berlin (ICCE-Berlin)}, pages 206--210, 2011.

\bibitem[Farke et~al.(2020)Farke, Lorenz, Schnitzler, Markert, and D\"{u}rmuth]{farke_2020}
Florian~M. Farke, Lennart Lorenz, Theodor Schnitzler, Philipp Markert, and Markus D\"{u}rmuth.
\newblock "you still use the password after all" — exploring fido2 security keys in a small company.
\newblock In \emph{Proceedings of the Sixteenth USENIX Conference on Usable Privacy and Security}, USA, 2020. USENIX Association.

\bibitem[Feth(2015)]{feth_2015}
Denis Feth.
\newblock User-centric security: Optimization of the security-usability trade-off.
\newblock In \emph{Proceedings of the 2015 10th Joint Meeting on Foundations of Software Engineering}, page 1034–1037, New York, NY, USA, 2015. Association for Computing Machinery.

\bibitem[Friedman and Bryen(2008)]{friedman_2008}
Mark~G. Friedman and Diane~Nelson Bryen.
\newblock Web accessibility design recommendations for people with cognitive disabilities.
\newblock \emph{Technology and Disability}, 19:\penalty0 205--212, 2008.

\bibitem[Gordon(2005)]{gordon_2005}
Wayne~A Gordon.
\newblock The interface between cognitive impairments and access to information technology.
\newblock \emph{ACCESSIBILITY AND COMPUTING}, page~3, 2005.

\bibitem[Hong et~al.(2010)Hong, Wang, Xu, Yan, and Chua]{hong_2010}
Richang Hong, Meng Wang, Mengdi Xu, Shuicheng Yan, and Tat-Seng Chua.
\newblock Dynamic captioning: video accessibility enhancement for hearing impairment.
\newblock \emph{Proceedings of the 18th ACM international conference on Multimedia}, 2010.

\bibitem[Hong et~al.(2011)Hong, Wang, Yuan, Xu, Jiang, Yan, and Chua]{hong_2011}
Richang Hong, Meng Wang, Xiao-Tong Yuan, Mengdi Xu, Jianguo Jiang, Shuicheng Yan, and Tat-Seng Chua.
\newblock Video accessibility enhancement for hearing-impaired users.
\newblock \emph{ACM Trans. Multim. Comput. Commun. Appl.}, 7:\penalty0 24, 2011.

\bibitem[Ismailova and Inal(2022)]{ismailova_2022}
Rita Ismailova and Yavuz Inal.
\newblock Comparison of online accessibility evaluation tools: an analysis of tool effectiveness.
\newblock \emph{IEEE Access}, 10:\penalty0 58233--58239, 2022.

\bibitem[Jain et~al.(2019)Jain, Desjardins, Findlater, and Froehlich]{jain_2019}
Dhruv Jain, Audrey Desjardins, Leah Findlater, and Jon~E. Froehlich.
\newblock Autoethnography of a hard of hearing traveler.
\newblock \emph{Proceedings of the 21st International ACM SIGACCESS Conference on Computers and Accessibility}, 2019.

\bibitem[Kaur et~al.(2022)Kaur, Kaur, and Shabaz]{kaur_2022}
Sandeep Kaur, Gaganpreet Kaur, and Mohammad Shabaz.
\newblock A secure two-factor authentication framework in cloud computing.
\newblock \emph{Security and Communication Networks}, 2022:\penalty0 1--9, 2022.

\bibitem[Khan and Akbar(2015)]{khan_2015}
Salman~H Khan and M~Ali Akbar.
\newblock Multi-factor authentication on cloud.
\newblock In \emph{2015 International Conference on Digital Image Computing: Techniques and Applications (DICTA)}, pages 1--7. IEEE, 2015.

\bibitem[Konoth et~al.(2020)Konoth, Fischer, Fokkink, Athanasopoulos, Razavi, and Bos]{konoth_2020}
Radhesh~Krishnan Konoth, Bj{\"o}rn Fischer, Wan~J. Fokkink, Elias Athanasopoulos, Kaveh Razavi, and Herbert Bos.
\newblock Securepay: Strengthening two-factor authentication for arbitrary transactions.
\newblock \emph{2020 IEEE European Symposium on Security and Privacy (EuroS\&P)}, pages 569--586, 2020.

\bibitem[Kraus et~al.(2022)Kraus, Durst, Ferreira, Veiga, Kailer, and Weinmann]{kraus_2022}
Sascha Kraus, Susanne Durst, Jo{\~a}o Jos{\'e}~Pinto Ferreira, Pedro~Mota Veiga, Norbert Kailer, and Alexandra Weinmann.
\newblock Digital transformation in business and management research: An overview of the current status quo.
\newblock \emph{Int. J. Inf. Manag.}, 63:\penalty0 102466, 2022.

\bibitem[Krombholz et~al.(2019)Krombholz, Busse, Pfeffer, Smith, and von Zezschwitz]{krombholz_2019}
Katharina Krombholz, Karoline Busse, Katharina Pfeffer, Matthew Smith, and Emanuel von Zezschwitz.
\newblock "if https were secure, i wouldn't need 2fa" - end user and administrator mental models of https.
\newblock \emph{2019 IEEE Symposium on Security and Privacy (SP)}, pages 246--263, 2019.

\bibitem[Kumar et~al.(2019)Kumar, Gao, Pirogova, and Fang]{kumar_2019}
Akshay Kumar, Lin Gao, Elena Pirogova, and Qiang Fang.
\newblock A review of error-related potential-based brain--computer interfaces for motor impaired people.
\newblock \emph{IEEE Access}, 7:\penalty0 142451--142466, 2019.

\bibitem[Leporini and Patern\`{o}(2004)]{leporini_2004}
Barbara Leporini and Fabio Patern\`{o}.
\newblock Increasing usability when interacting through screen readers.
\newblock \emph{Universal Access in the Information Society}, 3\penalty0 (1):\penalty0 57–70, 2004.

\bibitem[Lewis(2011)]{lewis_2011}
Clayton~H. Lewis.
\newblock Issues in web presentation for cognitive accessibility.
\newblock In \emph{Interacci{\'o}n}, 2011.

\bibitem[Lin et~al.(2020)Lin, Ilia, and Polakis]{lin_2020}
Xu Lin, Panagiotis Ilia, and Jason Polakis.
\newblock Fill in the blanks: Empirical analysis of the privacy threats of browser form autofill.
\newblock \emph{Proceedings of the 2020 ACM SIGSAC Conference on Computer and Communications Security}, 2020.

\bibitem[Liu et~al.(2015)Liu, Au, Huang, Lu, and Li]{liu_2015}
Joseph~K Liu, Man~Ho Au, Xinyi Huang, Rongxing Lu, and Jin Li.
\newblock Fine-grained two-factor access control for web-based cloud computing services.
\newblock \emph{IEEE Transactions on Information Forensics and Security}, 11\penalty0 (3):\penalty0 484--497, 2015.

\bibitem[Liu et~al.(2023)Liu, Zhou, Cao, Xu, Wang, Gao, Zeng, and Xu]{liu_2023}
Kaijun Liu, Zhou Zhou, Qiang Cao, Guosheng Xu, Chenyu Wang, Yuan Gao, Weikai Zeng, and Guoai Xu.
\newblock A robust and effective two-factor authentication (2fa) protocol based on ecc for mobile computing.
\newblock \emph{Applied Sciences}, 2023.

\bibitem[Moreno et~al.(2021)Moreno, Alarc{\'o}n, and Mart{\'i}nez]{moreno_2021}
Lourdes Moreno, Rodrigo Alarc{\'o}n, and Paloma Mart{\'i}nez.
\newblock Designing and evaluating a user interface for people with cognitive disabilities.
\newblock \emph{Proceedings of the XXI International Conference on Human Computer Interaction}, 2021.

\bibitem[Moreno et~al.(2023)Moreno, Petrie, Mart{\'\i}nez, and Alarcon]{moreno_2023}
Lourdes Moreno, Helen Petrie, Paloma Mart{\'\i}nez, and Rodrigo Alarcon.
\newblock Designing user interfaces for content simplification aimed at people with cognitive impairments.
\newblock \emph{Universal Access in the Information Society}, pages 1--19, 2023.

\bibitem[Mostafa et~al.(2023)Mostafa, Hassan, and Said]{mostafa_2023}
Elsayed Mostafa, MM Hassan, and Wael Said.
\newblock An interactive multi-factor user authentication framework in cloud computing.
\newblock \emph{International Journal of Computer Science \& Network Security}, 23\penalty0 (8):\penalty0 63--76, 2023.

\bibitem[Naftali and Findlater(2014)]{naftali_2014}
Maia Naftali and Leah Findlater.
\newblock Accessibility in context: Understanding the truly mobile experience of smartphone users with motor impairments.
\newblock In \emph{Proceedings of the 16th International ACM SIGACCESS Conference on Computers \& Accessibility}, page 209–216, New York, NY, USA, 2014. Association for Computing Machinery.

\bibitem[Nanayakkara et~al.(2013)Nanayakkara, Wyse, Ong, and Taylor]{nanayakkara_2013}
Suranga~Chandima Nanayakkara, Lonce Wyse, SH Ong, and Elizabeth~A Taylor.
\newblock Enhancing musical experience for the hearing-impaired using visual and haptic displays.
\newblock \emph{HUMAN--COMPUTER INTERACTION}, 28:\penalty0 115--160, 2013.

\bibitem[Neupert et~al.(2021)Neupert, Growney, Zhu, Sorensen, Smith, and Hannig]{neupert_2021}
Shevaun~D Neupert, Claire~M Growney, Xianghe Zhu, Julia~K Sorensen, Emily~L Smith, and Jan Hannig.
\newblock Bff: bayesian, fiducial, and frequentist analysis of cognitive engagement among cognitively impaired older adults.
\newblock \emph{Entropy}, 23\penalty0 (4):\penalty0 428, 2021.

\bibitem[Oren and Arad(2022)]{oren_2022}
Yossi Oren and D. Arad.
\newblock Toward usable and accessible two-factor authentication based on the piezo-gyro channel.
\newblock \emph{IEEE Access}, 10:\penalty0 19551--19557, 2022.

\bibitem[Pais et~al.(2020)Pais, Ruano, Carvalho, and Barros]{pais_2020}
Ricardo Pais, Lu{\'i}s Ruano, Of{\'e}lia~P Carvalho, and Henrique Barros.
\newblock Global cognitive impairment prevalence and incidence in community dwelling older adults—a systematic review.
\newblock \emph{Geriatrics}, 5, 2020.

\bibitem[Pascolini and Mariotti(2012)]{pascolini_2012}
Donatella Pascolini and Silvio~Paolo Mariotti.
\newblock Global estimates of visual impairment: 2010.
\newblock \emph{British Journal of Ophthalmology}, 96\penalty0 (5):\penalty0 614--618, 2012.

\bibitem[Pichiliani and Pizzolato(2021)]{britto_2021}
Talita Cristina Pagani~Britto Pichiliani and Ednaldo~Brigante Pizzolato.
\newblock Cognitive disabilities and web accessibility: a survey into the brazilian web development community.
\newblock \emph{J. Interact. Syst.}, 12:\penalty0 308--327, 2021.

\bibitem[Poulson and Nicolle(2004)]{poulson_2004}
David Poulson and Colette Nicolle.
\newblock Making the internet accessible for people with cognitive and communication impairments.
\newblock \emph{Universal Access in the Information Society}, 3:\penalty0 48--56, 2004.

\bibitem[Rajavenkatanarayanan et~al.(2019)Rajavenkatanarayanan, Kanal, Tsiakas, Calderon, Papakostas, Abujelala, Galib, Ford, Wylie, and Makedon]{rajavenkatanarayanan_2019}
Akilesh Rajavenkatanarayanan, Varun Kanal, Konstantinos Tsiakas, Diane Calderon, Michalis Papakostas, Maher Abujelala, Marnim Galib, James~C Ford, Glenn Wylie, and Fillia Makedon.
\newblock A survey of assistive technologies for assessment and rehabilitation of motor impairments in multiple sclerosis.
\newblock \emph{Multimodal Technologies and Interaction}, 3\penalty0 (1):\penalty0 6, 2019.

\bibitem[Reese et~al.(2019)Reese, Smith, Dutson, Armknecht, Cameron, and Seamons]{reese_2019}
Ken Reese, Trevor Smith, Jonathan Dutson, Jonathan Armknecht, Jacob Cameron, and Kent~E. Seamons.
\newblock A usability study of five two-factor authentication methods.
\newblock In \emph{SOUPS @ USENIX Security Symposium}, 2019.

\bibitem[Schneier(2005)]{schneier_2005}
Bruce Schneier.
\newblock Two-factor authentication: too little, too late.
\newblock \emph{Commun. ACM}, 48:\penalty0 136, 2005.

\bibitem[Scullion(2021)]{scullion_2021}
Dwyer Scullion.
\newblock Accessibility in academic publishing: more than just compliance.
\newblock Oxford University Press's Academic Insights for the Thinking World, 2021.
\newblock {https://blog.oup.com/2021/05/accessibility-in-academic-publishing-more-than-just-compliance/} {Accessed: 2023-11-24}.

\bibitem[Senjam et~al.(2021)Senjam, Manna, and Bascaran]{senjam_2021}
Suraj~Singh Senjam, Souvik Manna, and Covadonga Bascaran.
\newblock Smartphones-based assistive technology: Accessibility features and apps for people with visual impairment, and its usage, challenges, and usability testing.
\newblock \emph{Clinical Optometry}, 13:\penalty0 311 -- 322, 2021.

\bibitem[Su et~al.(2023)Su, Jiang, Du, Chen, Liu, Ren, Dai, Wang, Li, and Chen]{su_2023}
Yuchen Su, Guoqing Jiang, Yicong Du, Yuefeng Chen, Hongbo Liu, Yanzhi Ren, Huan Dai, Yan Wang, Shuai Li, and Yingying Chen.
\newblock P 2 auth: Two-factor authentication leveraging pin and keystroke-induced ppg measurements.
\newblock In \emph{2023 IEEE 43rd International Conference on Distributed Computing Systems (ICDCS)}, pages 726--737. IEEE, 2023.

\bibitem[Trewin et~al.(2009)Trewin, Laff, Hanson, and Cavender]{trewin_2009}
Shari Trewin, Mark Laff, Vicki Hanson, and Anna Cavender.
\newblock Exploring visual and motor accessibility in navigating a virtual world.
\newblock \emph{ACM Trans. Access. Comput.}, 2\penalty0 (2), 2009.

\bibitem[Ward and Shirley(2019)]{ward_2019}
Lauren Ward and Ben~G Shirley.
\newblock Personalization in object-based audio for accessibility: A review of advancements for hearing impaired listeners.
\newblock \emph{Journal of the Audio Engineering Society}, 67\penalty0 (7/8), 2019.

\bibitem[{World Health Organization (WHO)}(2023{\natexlab{a}})]{who_hi_2023}
{World Health Organization (WHO)}.
\newblock Deafness and hearing loss, 2023{\natexlab{a}}.
\newblock {https://www.who.int/news-room/fact-sheets/detail/deafness-and-hearing-loss/} {Accessed: 2023-12-02}.

\bibitem[{World Health Organization (WHO)}(2023{\natexlab{b}})]{who_mi_2023}
{World Health Organization (WHO)}.
\newblock Disability, 2023{\natexlab{b}}.
\newblock {https://www.who.int/news-room/fact-sheets/detail/deafness-and-hearing-loss/} {Accessed: 2023-12-09}.

\bibitem[{World Wide Web Consortium (W3C)}(2008)]{wcag_2008}
{World Wide Web Consortium (W3C)}.
\newblock Web content accessibility guidelines (wcag) 2.0, 2008.
\newblock ISO/IEC 40500:2012 standard.

\bibitem[{World Wide Web Consortium (W3C)}(2017)]{waiaria_2017}
{World Wide Web Consortium (W3C)}.
\newblock Accessible rich internet applications (wai-aria) 1.1, 2017.

\bibitem[{World Wide Web Consortium (W3C)}(2018)]{wcag_2018}
{World Wide Web Consortium (W3C)}.
\newblock Web content accessibility guidelines (wcag) 2.1, 2018.

\bibitem[{World Wide Web Consortium (W3C)}(2021)]{webauthn_2021}
{World Wide Web Consortium (W3C)}.
\newblock Web authentication: An api for accessing public key credentials level 2, 2021.

\bibitem[{World Wide Web Consortium (W3C)}(2023)]{wcag_2023}
{World Wide Web Consortium (W3C)}.
\newblock Web content accessibility guidelines (wcag) 2.2, 2023.

\bibitem[Zezulak et~al.(2023)Zezulak, Tazi, and Das]{zezulak_2023}
Alisa Zezulak, Faiza Tazi, and Sanchari Das.
\newblock Sok: Evaluating privacy and security concerns of using web services for the disabled population.
\newblock \emph{arXiv preprint arXiv:2302.13261}, 2023.

\bibitem[Zhu et~al.(2022)Zhu, Zhang, Zhang, Clepper, Jia, and Liu]{zhu_2022}
Yancong Zhu, Juan Zhang, Zhaoxi Zhang, Gina Clepper, Jingpeng Jia, and Wei Liu.
\newblock Designing an interactive communication assistance system for hearing-impaired college students based on gesture recognition and representation.
\newblock \emph{Future Internet}, 14\penalty0 (7):\penalty0 198, 2022.

\end{thebibliography}
}


\end{document}